\documentstyle[aps]
              {revtex}
\begin{document}
\draft
\title{
QND and higher order effects for a nonlinear meter
in an interferometric  gravitational wave antenna
}
\author{Yu. Levin
}
\address{Theoretical Astrophysics, California Institute
of Technology, Pasadena, California 91125}
\date{\today}
\maketitle
\begin{abstract}
A new optical topology and signal readout strategy for a laser interferometer
gravitational wave detector were proposed
recently by Braginsky and Khalili . Their method is based on using 
a nonlinear medium inside  a microwave  oscillator to detect the 
gravitational-wave-induced spatial shift
of the interferometer's standing optical wave. This paper proposes a 
quantum nondemolition (QND) scheme
that could be realistically used for such a readout device and discusses
a ``fundamental" sensitivity limit  imposed by a higher order optical effect.

\end{abstract}
\pacs{PACS numbers: xxxxxxxxxxx}

\narrowtext

\section{Introduction and Summary}

Laser interferometer gravitational wave detectors \\
(LIGO, VIRGO, GEO 600, TAMA)
are designed to detect small perturbations $h$ in the spatial metric
due to gravitational waves (GW) passing through the \\ Earth 
\cite{ligovirgo} . Being very  far from major astrophysical 
sources\cite{300yrs},  these detectors 
are likely to encounter GW's that are very weak, so the detectors must
be correspondingly sensitive---e.g., the first 
LIGO interferometer will be able to detect GW's with $h\sim 3\times 10^{-21}$ in the
frequency band of $30-300{\rm Hz}$. Improving the sensitivity of 
measurement may be necessary to achieve the first GW detection
and will surely be necessary to improve the event rate.

One of the major noise sources in traditional interferometers
 is the so-called shot noise. What is being detected
is the phase shift of the output optical wave\cite{ligovirgo}:
\begin{equation}
\delta\phi\sim\omega_{\rm opt}\tau_{\rm GW} h
\end{equation}
where $\omega_{\rm opt}$ is the angular frequency of the optical wave and $\tau_{\rm GW}$
is the half-period of a gravitational wave. For  coherent optical
pumping the uncertainty in the phase due to shot noise
is given by $\Delta \phi=1/\sqrt{N_{\rm GW}}$, where $N_{\rm GW}$ is the number of photons
introduced into the interferometer during $\tau_{\rm GW}$.
Thus a gravitational wave can be detected if 
\begin{equation}
N_{\rm GW}>N_{\rm min}\simeq {1\over (h\omega_{\rm opt}\tau_{\rm GW}^{*})^2}  .
\end{equation}
Therefore, in order to increase the gravity--wave sensitivity of the interferometer,
 we have to increase
the number of photons in the resonator (and hence the consumed laser
power) as $N\propto 1/h^2$. On the other hand, the presence of the large number
of optical photons in the resonator poses severe technical and 
fundamental \\ problems . Among the technical problems are distortion of
mirrors due to overheating,
and large laser power consumption\cite{ligovirgo}. The fundamental problem is that
photons in the interferometer will randomly buffet the mirrors inducing 
random motion indistinguishable from the motion produced by a gravitational wave.
 Balancing this radiation pressure noise and the shot noise
produces the Standard Quantum Limit (SQL) for monitoring the displacements
of the test masses\cite{300yrs}.

Recently Braginsky and Khalili have proposed a new way to improve the
sensitivity of an interferometric GW detector without increasing
the interferometer's optical power
\cite{BKmeter}.
Their method entails a new
type of GW readout based on 
a microwave oscillator containing an optically nonlinear
medium, which is placed inside the GW detector's high quality  Fabry-Perot resonator.
The advantage of this readout 
method is that,
unlike conventional interferometers, it does not require large optical
power circulating inside the FP resonator in order to achieve high sensitivity. 
In section $2$ the principles of this scheme are briefly outlined and
some numerical estimates are quoted.

Section 3 describes
a 
potentially practical Quantum Nondemolition (QND) strategy which can be 
used in the Braginsky--Khalili readout system (BK-meter). 
We show in Sec.3 and Appendix A [cf. Eq. (\ref{eq:star})]
that a QND measurement can be performed within a 
narrow frequency band centered around 
\begin{equation}
\Omega_0=\sqrt{6\hbar N\over m c L}\omega_{\rm opt},
\label{eq:star1}
\end{equation}
where $\omega_{\rm opt}$ and $N$ are the frequency of light and the number of photons
stored in the FP resonator respectively, $L$ is the distance between the end mirrors of
the FP resonator and $m$ is the mass of each of the test masses to which the mirrors are attached.
For $N=2.8\times 10^{20}$, $L=4\hbox{\rm km}$, $m=10 \hbox{\rm kg}$, $\omega_{\rm opt}=3\times 10^{15}$
one obtains $\Omega_0/2\pi=60\hbox{\rm Hz}$, which is within LIGO band. For the
resonator's relaxation time of 10 seconds (as assumed in \cite{BKmeter})
the necessary laser power to achieve this number of photons inside the
resonator is  $\sim 9\hbox{\rm Watt}$.

We  demonstrate in Appendix B that the bandwidth $\Delta \Omega$ of this measurement
determines the optimal power input to the microwave oscillator :
\begin{equation}
W_{\rm optimal}\sim W_{\rm SQL}{\Omega_0\over \Delta\Omega},
\label{eq:pumpower1}
\end{equation}
where $W_{\rm SQL }$ is the power input necessary to achieve 
the Standard Quantum Limit sensitivity at frequency $\Omega_0$;
cf.  Eq. (\ref{eq:pumpower}) where the expression and the
numerical estimate for $W_{\rm SQL}$ are given.
The signal-to-noise ratio achieved  by this QND measurement is
greater by a factor of
$\sqrt{{\Omega_0 /\Delta\Omega}}$ than the SQL:
\begin{equation}
\left({S\over N}\right)_{\rm QND}\sim\sqrt{{\Omega_0\over \Delta\Omega}}\left({S\over N}\right)_{\rm SQL} .
\end{equation}

  Section $4$ and Appedix B discuss a higher-order optical effect in the 
BK readout system 
and derive the sensitivity limit that it imposes . 
In particular, thermally excited mechanical modes
in the test masses will, after interacting with light
inside the FP resonator, produce a ``double conversion''
of photons, which will be registered as noise by the 
detector; cf. Eq. (\ref{eq:noisec}) and Eq. (\ref{eq:noisec1}).

\section{Principle of operation of the BK meter}

The layout of the BK meter is shown in Fig. 1
(for more detail the reader is referred to \cite{BKmeter}).
Three freely suspended mirrors---A,B and C---form walls of
an L-shaped  Fabry-Perot
(FP) resonator 
which supports a standing optical wave, driven by a laser
at end A or C. Section A-B of the resonator
would be in one arm of the LIGO (or other) vacuum system, and B-C in the other.
 The block D containing two thin slabs of non-linear
medium (Fig.2) is sandwiched between two thin focusing lenses two focal
lengths apart. The lenses and the block are attached to mirror B.

When the polarization tensor of a gravitational wave is aligned with the 
arms of the FP resonator, the distances between A and B and between
B and C will change in counterphase--i.e. when one is increasing,
the other one will decrease. This will produce the a net spatial shift
of the standing optical wave with respect to mirror B, thus changing
the amplitude of the optical field within the two slabs of
nonlinear medium. 
The slabs have cubic nonlinearities that are equal in magnitude but opposite in
sign. They are positioned  
symmetrically with respect to the crest
of the standing  optical wave as shown in Fig.2.
Block D, which contains the slabs, is placed in between the plates of a capacitor which
in turn is part of a microwave oscillator. 

The spatial shift in the 
optical standing wave produces
changes of electric field in the first and second slabs that are equal in
magnitude and opposite in sign: 
\begin{equation}
\delta E_1 =-\delta E_2 .
\label{eq:spshift}
\end{equation}
Since the two  slabs have the opposite
nonlinearities, $\chi_1^{(3)}=-\chi_2^{(3)}$, the change in the index of 
refraction is the same for both of them:
\begin{equation}
\delta n_1 = \delta n_2 = 4\pi \chi_1^{(3)} E_1 \delta E_1.
\end{equation}
This change in dielectric constants of the plates  in turn changes the value
of the microwave oscillator's capacitance, thus producing a shift in its resonant frequency:  

\begin{equation}
\delta \omega_{\rm e}={K \omega_{\rm opt}\over 2} h
\label{eq:microshift}
\end{equation}
where $\omega_{e}$ is the frequency of the microwave oscillator, $\omega_{\rm opt}$ is the frequency
of the optical wave, and \\ 
$K=16\pi^2\chi^{(3)}lN\hbar\omega_{\rm opt}\omega_e/Vc$. Here $l$ is the
width of each of the nonlinear slabs and $V$ is the volume of the capacitor.
This shift is seen as a phase shift in the readout of the microwave
oscillator:
\begin{equation}
\delta \phi=\delta \omega_{\rm e} \tau_{\rm e}^{*}={1\over 2}K\omega_{\rm opt}\tau_{\rm e}^{*}h ,
\label{eq:phaseshift}
\end{equation}
where $\tau_{\rm e}^{*} $ is the oscillator's ringdown time.  

Braginsky and Khalili compare this with the traditional
optical readout schemes in which the phase shift of the 
recombined optical wave is detected:
\begin{equation}
\delta \phi_{\rm opt}=\omega_{\rm opt}\tau_o^{*} h,
\label{eq:optphase}
\end{equation}
where $\tau_o^{*}$ is the  ringdown time of the two traditional FP resonators,
one in each arm of the interferometer.
For $\chi^{(3)}=10^{-14} {\rm cm}^2/{\rm Volt}^2$  (fused silica)
, $E_{\rm opt}^2 = 10^7 {\rm Volt}^2/{\rm cm}^2 $ (optical breakdown of fused
silica) they calculate $K$ in  Eq. (\ref{eq:microshift})  to be of
order $1$, so for $\tau_{\rm opt}^{*}\sim\tau_{\rm e}^{*}$ the responses
of both systems in terms of phaseshift are of the same order,
\begin{equation}
\phi_{\rm opt} \sim \phi_{\rm e}.
\end{equation}
For coherent pumping in both cases the uncertainty in the phase is 
$\Delta\phi\sim 1/\sqrt{N_{\rm GW}}$, where $N_{\rm GW}$ is the number of photons 
(optical in conventional interferometers and microwave in the BK
readout system) introduced into the interferometer during
an averaging time (half the GW period). 
So to achieve the same level of sensitivity one needs to pump the
same number of photons in both cases, but the power needed by the 
BK meter is smaller by a factor of $\omega_{\rm opt}/\omega_{\rm e}\sim 10^4$.
The BK estimate for the microwave power is 
\begin{equation}
W_{\rm e}={\hbar \omega_{\rm e} N_{\rm_e}\over \tau_{\rm e}^{*}} \sim
                   1\hbox{Watt}
\end{equation}
for $N_{\rm e}\sim 10^{20}$. 
For more detailed estimates the reader is refered to Ref.\cite{BKmeter} .

\section{QND for the BK readout system}

Any readout system that monitors the displacement of the mirrors
must exert on them a fluctuating back action force,
thus enforcing the Heisenberg uncertainty relation .
As a consequence of this, all straightforward displacement measurements
run into the Standard Quantum Limit (SQL) \cite{QND}, \cite{QND1}
\begin{equation}
\Delta x_{\rm SQL}=\sqrt{\hbar\tau\over m},
\label{eq:sqls}
\end{equation}
where $\Delta x$ is the minimal uncertainty in displacement of a free mass $m$
monitored over a time interval $\tau$. This SQL for displacement 
can also be written
in terms of the limiting spectral density of the mirrors'
displacement fluctuations\cite{QND}:
\begin{equation}
S_{x}^{\rm SQL}(\Omega)={\hbar\over m\Omega^2}
\end{equation}
where $\Omega$ is the  frequency. Then $\Delta x=\sqrt{S_x^{\rm SQL}\Delta\Omega}$,
and for $\Omega\sim\Delta\Omega\sim 1/\tau$ one recovers Eq. (\ref{eq:sqls})

The SQL for a free mass
is by no means a fundamental limit; it can be overcome by a variety 
of techniques  \cite{prevmethods} which are known collectively 
as Quantum Nondemolition
(QND) measurements. All previously proposed QND schemes that are applicable for conventional 
GW interferometers utilize highly non-classical states of light, and none of them
are practical because of technical difficulties (most especially because
of the large required optical pumping power and because losses so easily destroy the 
non-classical states of light). In this paper a different 
strategy is proposed, one which does not require the deliberate creation or detection
of any non-classical state of light and thus can be more practically implemented.
This scheme, however, is confined to narrow-band measurements.

We begin by describing the backaction mechanism by which the BK readout system
enforces the Heisenberg Uncertainty relation on the measurement of the test-mass
position.
The quantum state of the BK microwave oscillator satisfies the usual phase-number
uncertainty relation
\begin{equation}
\Delta\phi_{\rm e}\Delta N_{\rm e}>{1\over 2}.
\end{equation}
The more accurately the BK-meter reads out $\phi_e$, the larger
will be the fluctuations $\Delta N_e$ in the oscillator's number
of microwave photons.
The $\chi^{(3)}$ nonlinearity will transform $\Delta N_{\rm e}$ into an uncertainty
of the optical index of refraction of the slabs:
\begin{equation}
\delta n_1=-\delta n_2 = {16\pi^2\chi^{(3)}\hbar\over\epsilon V}\delta N_{\rm e},
\label{eq:refrindex}
\end{equation}
where $\delta N_{\rm e}$ is the fluctuation in $N_{\rm e}$,  $\delta n_1$ and
 $\delta n_2$ are the resulting fluctuations in $n_1$ and $n_2$, $\epsilon$ is the
coefficient
of dielectric permitivity and $V$ is the volume of the capacitor.
Braginsky and Khalili have argued\cite{BKmeter}
 that $\delta n_1$ and $\delta n_2$ cause a
redistribution  of the optical energy between the left and the right parts
of the FP resonator,  thereby giving rise to  a net difference in the 
forces buffeting the mirrors: 
\begin{equation}
\delta F=K{\omega_{\rm opt}\over\omega_{\rm e}}{\hbar\delta N_{\rm e}\over L},
\end{equation}
where $L$ is the total length of the FP resonator.
This fluctuating  force will cause fluctuations in the positions of 
the mirrors, thus causing fluctuations in the spatial shift of the optical
field with respect to the mirror B and the nonlinear slabs attached to it:
\begin{equation}
\delta\tilde x(\Omega)=-{3\over 2}{\delta\tilde F(\Omega)\over m \Omega^2}=
  -{3\over 2} K{\omega_{\rm opt}\over\omega_{\rm e}}{\hbar\over L}\delta
   \tilde N_{\rm e}(\Omega),
\label{eq:backact}
\end{equation}
where $\delta\tilde x(\Omega)$, $\delta\tilde F(\Omega)$
 and $\delta\tilde N_e(\Omega)$ are Fourier components of the corresponding
quantities, $m$ is the mass of each of the mirrors , and the  factor ${3\over 2}$
comes about when motion of all three mirrors is taken into account.

Now we are ready to describe our QND method, but first  the following simple
remark must be made. Suppose for a moment that all of the mirrors
are rigidly fixed. As already mentioned above, fluctuations in $N_{\rm e}$,
by changing the optical coefficient of refraction of the slabs [Eq. (\ref{eq:refrindex})], will
redistribute optical energy between the left and right parts of the FP resonator.
A straightforward calculation [Eq. (\ref{eq:deltaE}) in Appendix A] shows that this alone
will change the optical field inside the nonlinear slabs so that
\begin{equation}
\delta E_1=-\delta E_2=E_0{n\omega_{\rm opt}l\over c\sqrt{2}}\delta n,
\end{equation}
thereby  simulating a spatial shift of the optical 
field as in Eq. (\ref{eq:spshift}). Here $l$ is the width of the nonlinear slab , $n\equiv n_1=n_2$ , 
$\delta n\equiv\delta n_1=-\delta n_2$ and $E_0$ is the peak amplitude of the
optical standing wave inside the FP resonator.

Now if we release the mirrors, the back action (\ref{eq:backact})
will affect our reading
as well, and the total fluctuations of the optical field inside the nonlinear
slabs will be given by
\begin{eqnarray}
\delta\tilde E_1(\Omega)&-&\delta\tilde E_2(\Omega)=\label{eq:16}\\
                        &=&\sqrt{2} n E_0{\omega_{\rm opt}l 
                          \over c}\left(1 -6{\omega_{\rm opt}^2\over
                          \Omega^2}{\hbar N\over m c^2\tau}\right)\delta n
                          (\Omega)\nonumber
\end{eqnarray}
where $\tau=L/c$; cf. Eq. (\ref{eq:deltaE1}) of Appendix A.
From the above equation we see that  for a given frequency $\Omega=\Omega_0$ we can
adjust $N$ in such a way that $\delta\tilde E_1(\Omega)=
\delta\tilde E_2(\Omega)=0$ and thus the readout system does not register  any fluctuations
due to the back action (but only for that value of $\Omega$). Thus a QND measurement
is performed. The relationship between the QND angular frequency 
$\Omega_0$ and the number $N$ of optical photons in the Fabry-Perot
resonator is
\begin{equation}
\Omega_0=\sqrt{6\hbar N\over m c L}\omega_{\rm opt};
\label{eq:star}
\end{equation}
see Eq. (\ref{eq:star1}) of Sec. I.

The essential reason that this readout is QND is that it registers  
not only the fluctuations of the mirrors' displacement $x$ due to backaction
[the second term in large parentheses in Eq. (17)], but also directly the
back-action force (the first term). Thus a position-momentum correlation is introduced into
the measurement procedure, and such correlations are known to make QND possible\cite{QND}.
For $L=4\hbox{km}$, $\Omega/2\pi=60\hbox{Hz}$, $\omega_{\rm opt}=
3\times10^{15}\hbox{s}^{-1}$, and
$m=10\hbox{kg}$ the necessary number of optical photons to perform QND is
\begin{equation}
N={1\over 6}{\Omega_0^2\over \omega_{\rm opt}^2}{mc^2\tau\over\hbar}\sim 2.8\times
10^{20}.
\label{eq:N}
\end{equation}
The QND measurement described above is clearly  narrowband .
But one can dynamically tune the frequency at which the QND  is
performed by changing the laser power and thus changing the number
of optical photons $N$ in the resonator, provided that the frequency of the
signal changes slowly compared to the ring-down rate of the optical
resonator. 

 Appendix B considers a particular scheme for measuring of the phase
of the microwave oscillator. In this scheme the oscillator
is coupled to a transmission line, and the physically measured quantity
is the phase quadrature of the outgoing electromagnetic wave 
propagating along the transmission line. Having specified fully
the measurement model, 
we find that if the bandwidth of measurement is $\Delta\Omega$
then the signal-to-noise ratio for the narrow-band QND measurement can be as high as
\begin{eqnarray}
\left({S\over N}\right)^2_{\rm QND}&\sim&\left({S\over N}\right)^2_{\rm SQL}{\Omega_0\over\Delta\Omega}\label{line1}\\
       & &={1\over2\pi}{\Omega_0\over\Delta\Omega}
                            \int_{\Omega_0-\Delta\Omega}^{\Omega_0+\Delta\Omega}{m\Omega^2 L^2 |\tilde{h(\Omega)}|^2\over\hbar}d\Omega\nonumber .
\end{eqnarray}
 The above  signal-to-noise ratio is achieved when the  pumping power
of the microwave oscillator is given by
\begin{eqnarray}
W_{\rm optimal}&=&{V^2\over 32\pi^2{\chi^{(3)}}^2N\tau}
\left({\Omega_0\over\omega_{\rm opt}}\right)^2 \left({L\over l}\right)^2
 {1\over\hbar\omega_e}{\Omega_0\over\Delta\Omega}\nonumber\\
 &=&W_{\rm SQL}{\Omega_0\over \Delta\Omega},\label{line1} 
\label{eq:pumpower}
\end{eqnarray}
where $V$ is the volume of the capacitor
and $W_{\rm SQL}$ is the minimal power necessary to achieve the 
SQL sensitivity level; cf. Eq. (\ref{eq:optpower}) of Appendix B and Eq. (\ref{eq:pumpower1})
of the introduction. For $V=(.01\hbox{mm})^3$, $\omega_e=10^{11} \hbox{s}^{-1}$ and for
other parameters having numerical values as in Eq. (\ref{eq:N}), we get
$W_{\rm optimal}=0.1\hbox{Watt} \left(\Omega_0/\Delta\Omega\right)$ (cf. the $1$kW of the optical
power required to achieve the SQL in a conventional interferometric scheme).

While we have not devised a general proof, it seems likely
that no other microwave readout scheme can operate with a power
less than in Eq. (\ref{eq:pumpower}); expression (\ref{eq:pumpower})
is probably a general relation for optimally disigned microwave 
readout schemes.

\section{Higher-order optical effects; fundamental sensitivity limit}

In this section we identify and discuss a fundamental limit on the sensitivity
of the BK readout system---a limit that applies whether or not the system is 
being operated in a QND mode.

In an interferometric GW detector,  mirrors are installed on the surfaces of 
 test masses, which have internal elastic mechanical modes of frequencies 
$\Omega_m/2\pi\ge 12 \hbox{ kHz}$. The noise curve of the interferometer will have large
peaks near these frequencies. When photons of frequency $\omega_{\rm opt}$ interact with
walls oscillating with the frequancy $\Omega_m$, some of the photons will be up or down
converted to frequencies $\omega_{\rm opt}\pm\Omega_m$. These up or down
converted photons in turn interact with the ``noisy'' walls, and if there is a non-zero
component of the mirrors' motion at $\Omega_m\pm\Omega$, then some of the photons will
up or down convert a second time to frequencies $\omega_0\pm\Omega$. If $\Omega$
is the frequency of detection, then this  second-order process 
of double frequency conversion will
be registered by the BK readout system as a signal from a gravitational wave.

The perturbation theory for FP resonators with moving walls is worked 
out in detail in Appendix C. Here just the main result is quoted.
From Eq. (\ref{eq:spdensity}) the noise curve in units of
$1/\sqrt{\rm Hz}$ is given by
\begin{equation}
\sqrt{S_h(\Omega)}\sim {1\over 3\pi}{\omega_{\rm opt} k_B T_e\sqrt{\gamma_m}\over
                                m   L\Omega_m^2\Omega^2\tau c}\sqrt{\Omega\tau+\epsilon}
\label{eq:noisec}
\end{equation}
where $T_e$ is the temperature of the test masses, $\gamma_m$ is the damping rate of their mechanical
modes, and $\epsilon\sim\max{(\delta\chi^{(3)}/ \chi^{(3)}, \delta l/ \lambda)}$.
Here $\delta\chi\equiv\chi^{(3)}_1+\chi^{(3)}_2$, $\chi^{(3)}\equiv\chi^{(3)}_1$, $\lambda$ is the wavelength of light in the resonator
and $\delta l$ is the spatial offset of the slabs  from the position in shown Fig.\ 2.
For $L=4\hbox{ km}$, $\Omega=60\hbox{rad/sec}$, $T_e=300\hbox{K}$, $\Omega_m=7.2\times 10^4 \hbox{rad/sec}$,
$m=10\hbox{kg}$ and $\gamma_m=10^{-8}\Omega_m$
we get the noise level of
\begin{equation}
\sqrt{S_h}\sim 10^{-28}\sqrt{(\Omega\tau+\epsilon)}/\sqrt{\hbox{Hz}}
\label{eq:noisec1}
\end{equation}

It is not  unimaginable that future interferometers will achieve  sensitivities
$\sqrt{S_h}\sim 10^{-29}/\sqrt{\rm Hz}$
for low frequency ($10-100\hbox{Hz}$) narrow band signals (by e.g. using the  QND technique 
described in this paper).
In this case, Eqs.\ (\ref{eq:noisec}) and (\ref{eq:noisec1}) show that higher order effects
 will 
give rise to a ``fundamental'' low frequency noise limit
of magnitude
\begin{equation}
\sqrt{S_h(\Omega)}\sim 10^{-28}\left({60\hbox{rad/s}\over\Omega}\right)^2/\sqrt{\hbox{Hz}}.
\end{equation}

\section{Conclusions}
In this paper we have shown that  a practical QND measurement
might be  possible for a narrow-band 
measurement by a gravitational wave interferometer
using a BK readout system. Also it was shown that  second-order 
effects
set a   ``fundumental'' limit  
on the precision of the measurement.

\section{Acknowledgements}
I want to thank Vladimir Braginsky for suggesting the problem,
for interesting and useful discussions, and for wonderful
Pasadena ice-cream,  Kip Thorne
for reading carefully over the manuscript
and making many useful suggestions, and Farid Khalili
and Sergei Vyatchanin for interesting discussions
and excellent tea. I also thank Glenn Sobermann
for his advice on the prose, and Michele Pickerell
and Ruben Krasnopol'skiy for their help with
computer graphics.
This work was supported in part by 
NSF grant PHY-9424337.

\begin{appendix}
\section{ physics of the nonlinear medium inside the BK Fabry-Perot resonator}

First consider one slab of nonlinear medium positioned inside a  FP resonator.
Let $x_1$ (Fig. 3) be the total path length from the left mirror to the left edge
of the slab, $x_3$ be the path length from the right mirror to the right 
edge of the slab and $l$  be the width of the slab. For simplicity of the calculation
we assume $l\ll \lambda$ where $\lambda$ is the wavelength of light in the
resonator. Also for convenience define  $\tau_1=x_1/c$, $\tau_2=l/c$,
$\tau_3=x_3/c$.

The eigenfrequencies $\omega$ of this optical resonator were worked out in
\cite{BKmeter}. They satisfy
the following eigenequation:
\begin{eqnarray}
\sin (\omega\tau)&=&(n-1)\sin (n\omega\tau_2)\nonumber\\
                 & &\left[
            \sin \left(\omega\tau_1\right)\sin\left(\omega\tau_3\right)+
                  {1\over n}\cos \left(\omega\tau_1\right) \cos\left(
                    \omega\tau_3\right)\right]\label{line1}
\end{eqnarray}
where $\tau=\tau_1+\tau_3+n\tau_2$ .
This equation has  approximate solutions
\begin{equation}
\omega=\omega_0+(n^2-1)\omega_0{\tau_2\over 2\tau}\left\{\cos\left[\omega_0
       \left(\tau_1-\tau_3
        \right)\right] -1 \right\}
\end{equation}
where $\omega_0=\pi k/\tau$, and $k$ is any integer. When the slab's index of refraction
$n$ changes, $\omega$ changes accordingly:
\begin{equation}
{d\omega\over d n}={n\omega^0\tau_2\over \tau}\left\{\cos\left[\omega^0\left(\tau_1-\tau_3\right)
\right]-1\right\}
\label{eq:resfreq}
\end{equation}
 
The total optical energy contained in the resonator is 
\begin{equation}
U=N\hbar\omega
\end{equation}
where $N$ is the number of optical photons . We can find all of the forces
acting on the mirrors by taking derivatives of $U$ with respect to $\tau_1$
and $\tau_3$. For example,
\begin{equation}
F_{\rm left}=-{N\hbar\over c}{\partial\omega\over\partial\tau_1}
\label{eq:force1}
\end{equation}
and
\begin{equation}
F_{\rm right}=-{N\hbar\over c}{\partial\omega\over\partial\tau_3}
\label{eq:force2}
\end{equation}
where $F_{\rm left}$ and $F_{\rm right}$ are the forces acting on
the left and the right mirrors respectively, with the positive direction being out of the 
resonator. When taking derivatives of $\omega$ one has to keep in mind that
$\omega_0$ also depends on $\tau_1$ and $\tau_3$. 

The force acting on the slab of nonlinear medium is $F_{\rm left}-F_{\rm right}$ . The total 
spatial shift 
of the optical wave with respect to the slab due to the forces acting
on  the end mirrors and the slab itself is
\begin{equation}
\delta\tilde x(\Omega)=-{3\over 2 m \Omega^2}\left[\tilde F_{\rm left}(\Omega)-
                        \tilde F_{\rm right}(\Omega)\right],
\label{eq:force}
\end{equation}
where ``tildas'' stand for Fourier Transforms. If $F_{\rm left}$ and 
$F_{\rm right}$ are produced by a fluctuating index of refraction  $n=n_0+\delta n$
then on substituting Eqs.\ (\ref{eq:force1}) and (\ref{eq:force2})
into Eq. (\ref{eq:force}) we get
\begin{equation}  
\delta\tilde x(\Omega)={3 N\hbar\over 2 m \Omega^2}\left({\partial\over
\partial\tau_1}-{\partial\over\partial\tau_3}\right){d\omega\over d n}
\delta\tilde n(\Omega).
\label{eq:tildx}
\end{equation}
By then putting Eqs.\ (\ref{eq:resfreq})  and (\ref{eq:tildx}) together we obtain
\begin{equation}
\delta\tilde x(\Omega)=-{3N\hbar n\over m c }{\omega^2\over\Omega^2}{\tau_2
\over\tau}\sin\left[\omega\left(\tau_1-\tau_3\right)\right]\delta\tilde n(\Omega).
\end{equation}

For the two slabs of  opposite nonlinearities ($\delta n_1=-\delta n_2 =\delta n$)
in the configuration of Fig.\ 2 their two contributions add up to give
\begin{equation}
\delta\tilde x(\Omega)=-6 {N n \hbar\over m c}{\omega^2\over\Omega^2}
{\tau_2\over\tau}\delta\tilde n(\Omega).
\label{eq:backaction}
\end{equation}
The above expression is a manifestation of the back action as explained
in Sec. II.

Now let the amplitude of the optical electric field in the left part of the resonator be 
\begin{equation}
E_{\rm left}=E_0\sin\left({\omega x\over c}\right)
\end{equation}
where $x$ is the spatial coordinate with the origin at the left wall.
Then the field in the middle of the left slab is given by 
\begin{eqnarray}
E_1&=&E_0\left[\sin\left(\omega\tau_1\right)+{1\over n}
\cos\left(\omega\tau_1\right)\sin\left({n\omega\tau_2\over 2}\right)\right]\nonumber\\
& &+
      O\left[\left(\omega\tau_2\right)^2\right].\label{line1}
\end{eqnarray}
Now
\begin{equation}
{d E_1\over d n}\simeq {d E_0\over d n}\sin(\omega\tau_1)+E_0\tau_1
\cos(\omega\tau_1){d \omega\over d n}.
\end{equation}
In the case when two slabs are present inside the FP resonator, 
their contributions to the frequency and
field changes add up linearly (since the perturbations are very small). 
For the
configuration of Fig. 2 we see from Eq.(\ref{eq:resfreq}) that $d \omega/dn =0$, so

\begin{equation}
 {d E_1\over d n}={d E_0\over d n}\sin (\omega\tau_1).
\label{eq:32}
\end{equation}
But $F_{\rm left}\propto E_0^2$, so
\begin{equation}
{d E_0/d n\over E_0}={1\over 2}{d F_{\rm left}/ d n\over F_{\rm left}}=
-{\tau\over \omega}{d\over d n}{\partial\omega\over\partial\tau_1}.
\label{eq:33}
\end{equation}
Putting Eq.(\ref{eq:resfreq}) and Eq.(\ref{eq:32}) into Eq.(\ref{eq:33})
 and doing exactly the same calculation for 
the second slab, we obtain

\begin{equation}
\delta E_1=-\delta E_2=E_0{n\omega_{\rm opt}l\over \sqrt{2} c}\delta n
\label{eq:deltaE}
\end{equation}
for the case when mirror B is in the middle of the resonator.
Combining this with the back action from Eq.(\ref{eq:backaction})
we finally get Eq.(\ref{eq:16}) of Sec. 3:
\begin{eqnarray}
\delta\tilde E_1(\Omega)&-&\delta\tilde E_2(\Omega)=\label{eq:deltaE1}\\
                        &=&\sqrt{2} n E_0{\omega_{\rm opt}l
                          \over c}\left(1 -6{\omega_{\rm opt}^2\over
                          \Omega^2}{\hbar N\over m c^2\tau}\right)\delta n
                          (\Omega).\nonumber
\end{eqnarray}.

\section{ Calculation of optimal microwave power and signal-to-noise ratio for a QND measurement}

Consider the  microwave oscillator as shown on Fig. 2. In order to get
information about the phase of the oscillator, we have to couple
it to the outside world. Whatever the nature of this coupling
is, it will cause dissipation of the induced oscillations and hence,
by the fluctuation-dissipation theorem, give birth to a fluctuating
component of the oscillator's current. 

For concreteness, we
 model this coupling by an open transmission
line of impedance $R$. We assume that the oscillator,
consisting of the capacitor $ C$ and inductor ${\cal L}$, 
is driven on resonance by a  generator G with a voltage
output of amplitude $V_0$ (see Fig. 2). We also assume the transmission
line encompasses all of the dissipation present in the oscillator,
i.e., more generally, that we can access all of the information
escaping from the oscillator. And finally, we set the temperature
of the outside world to $0$ (in reality, one will  have to 
cool the oscillator to temperatures below the ones corresponding
to a microwave frequency).
The ingoing vacuum modes  drive fluctuations
in the circuit as described above, and the phase of the outgoing wave
contains information about the phase of the oscillator.

 The ingoing
modes are described by the  positive frequency part of a voltage operator 
\begin{equation}
V_{\rm in}=\int_0^\infty d\omega\sqrt{R\hbar\omega} a_{\rm in}(\omega)
               e^{-\imath \omega t} ,
\label{eq:voltage}
\end{equation}
where $a_{\rm in}(\omega) $ is the annihilation  operator for the
ingoing mode of frequency $\omega$ normalized so that
$\langle0|a_{\rm in}(\omega) a^{\dagger}_{\rm in}(\omega^\prime)|0\rangle =\delta (\omega-
                                                  \omega^\prime) $. Then the
fourier component of the outgoing wave is
\begin{eqnarray}
V_{\rm out}(\omega_e+\Omega)&=&\sqrt{R\hbar\omega_e}
                               {\alpha+\imath\Omega\over\alpha-\imath\Omega}
                               a_{\rm in}(\omega_e+\Omega)\nonumber\\
                            &+&{V_0\delta\omega_e(\Omega)\over
                                  2\Omega+\imath\alpha},\label{line1}
\label{eq:vout}
\end{eqnarray}
where $\alpha=R/{\cal L}$ is the ringdown rate of the microwave oscillator
and $\delta\omega_e$ is the variation in the oscillator's resonant frequency
due to fluctuating optical fields in the slabs of nonlinear medium, as explained
in Sec. II:

\begin{equation}
\delta\omega_e={8\pi^2\chi^{(3)}l N\hbar\omega_{\rm opt}\omega_e\over
                \sqrt{2}VL}{\delta E_1-\delta E_2\over E_0}.
\label{eq:deltaomegae}
\end{equation}
Here $V$ is the volume of the capacitor. The change of the optical field
inside the slabs is given by 
\begin{eqnarray}
\delta E_1 (\Omega)-\delta E_2(\Omega)&=&\sqrt{2}E_0 n\omega{\rm opt}\tau_2
                               \left(1-6{N\hbar\over mc^2\tau}{\omega_{\rm
                               opt}^2\over \Omega^2}\right)\delta n (\Omega)\nonumber\\
                           & &+\sqrt{2}E_0{\omega_{\rm opt}\over c} x_{\rm s}(\Omega),\label{line1}
\label{eq:deltae}
\end{eqnarray}
where the first term on the right hand side is due to the fluctuating index of refraction
of the slabs [cf. Eq. (\ref{eq:16}) of  Sec. III  and discussion therein], and the second
term is due to the GW-induced relative displacement $x_s$ of the slabs with respect to the standing
optical wave.
The fluctuations $\delta n$ of the indices of refraction of the nonlinear slabs  in the
above expression are caused by the voltage fluctuations on the plates of the capacitor,
which in turn can be traced to the incoming vacuum modes of the transmission line:
\begin{equation}
\delta n(\Omega)=-\imath{2\pi\chi^{(3)}\sqrt{R\hbar\omega_e}V_0\omega_e^2\over
                  \alpha (2\Omega+\imath\alpha)d^2}\left[a_{\rm in}(\omega_e+\Omega)+
                    a^{\dagger}_{\rm in}(\omega_e-\Omega)\right],
\label{eq:deltan}
\end{equation}
where $d$ is the distance between the plates of the capacitor.
Collecting Equations (\ref{eq:vout}), (\ref{eq:deltaomegae}), (\ref{eq:deltae})
and (\ref{eq:deltan}) together, we can write down the expression 
for the phase quadrature of the outgoing wave in the transmission line,
which is the measured readout signal:
\begin{eqnarray}
r(\Omega)&=&\left[V_{\rm out}(\omega_e+\Omega)-V_{\rm out}^{\dagger}(\omega_e-\Omega)\right]
                      /V_0\nonumber\\                
         &=&       {\sqrt{R\hbar\omega_e}\over V_0}{\alpha+\imath\Omega\over\alpha-\imath\Omega}
               \left[a_{\rm in}\left(\omega_e+\Omega\right)-a^{\dagger}_{\rm in}\left(\omega_e-\Omega\right)
                 \right]+\label{line1}\\
         &+& {8\pi^2 V_0\chi^{(3)}l N \hbar \omega_{\rm opt}^2\omega_e\over V_0(2\Omega+\imath\alpha)VLc}
             \left\{x_s-\right.\nonumber\\
         & &\left.-{8\pi^2\chi^{(3)} V_0\sqrt{R\hbar\omega_e}l\over R V (2\Omega+\imath\alpha)}\left(
                1-{6N\hbar\over mc^2\tau}{\omega_{\rm opt}^2\over\Omega^2}\right)\left[a_{\rm in}\left(
                 \omega_e+\Omega\right)+a_{\rm in}^{\dagger}\left(\omega_e-\Omega\right)\right]\right\}.\nonumber
\label{eq:r(\Omega)}
\end{eqnarray}
The measured $x$ is then given by 
\begin{eqnarray}
x_{\rm measured}(\Omega)&=&x_s(\Omega)-{8\pi^2\chi^{(3)}\sqrt{\hbar\omega_e W}l\over V (2\Omega+\imath
                            \alpha)}\left(1-{6N\hbar\over m c^2\tau}{\omega_{\rm opt}^2\over \Omega^2}
                           \right)\left[a_{\rm in}\left(\omega_e+\Omega\right)+a^{\dagger}_{\rm in}\left(
                           \omega_e-\Omega\right)\right]+\nonumber\\
                        & &+{VL c(2\Omega+\imath\alpha)\sqrt{\hbar\omega_e}\over 16\pi^2 \chi^{(3)}l N
                           \hbar\omega_{\rm opt}^2\omega_e\sqrt{W}}\left[a_{\rm in}\left(\omega_e+\Omega\right)-
                          a_{\rm in}^{\dagger}\left(\omega_e-\Omega\right)\right],\label{line1}
\label{eq:xmeasured}
\end{eqnarray}
where $W=V_0^2/R$ is the power pumped into the microwave oscillator by the generator G. The 
corresponding spectral density
of the Gaussian noise seen by the readout system is
\begin{eqnarray}
S_x(\Omega)&=&\left({8\pi^2\chi^{(3)}l\over V}\right)^2{\hbar\omega_e W\over 4\Omega^2+\alpha^2}
               \left(1-{6N\hbar\over m c^2 \tau}{\omega_{\rm opt}^2\over \Omega^2}\right)^2+\nonumber\\
           &+&\left(VLc\over 16\pi^2\chi^{(3)}l N\hbar\omega_{\rm opt}^2\omega_e\right)^2
                {(4\Omega^2+\alpha^2)\hbar\omega_e\over W}.\label{line1}
\label{eq:spdens}
\end{eqnarray} 
The first term on the right hand side corresponds to the back-action noise and the 
second term corresponds to the intrinsic noise of the measuring device.

 We
aim to perform a measurement with a narrow frequency band centered 
around the frequency $\Omega_0$ at which the back-action noise is zero:
\begin{equation}
\Omega_0=\sqrt{6N\hbar\over m\tau}{\omega_{\rm opt}\over c}.
\label{eq:Omega0}
\end{equation}
We write $S_x$ as a Taylor expansion in frequency around $\Omega_0$ :
\begin{equation}
S_x(\Omega)\simeq A(\Omega_0)W{(\Omega-\Omega_0)^2\over\Omega_0^2}+{B(\Omega_0)\over W},
\label{eq:spdens1}
\end{equation}
where $A$ and $B$ can be read from Eq. (\ref{eq:spdens}).  If the relevant bandwidth is $\Delta\Omega$
then we place a limit $(\Omega-\Omega_0)^2\leq\Delta\Omega^2 $ and 
\begin{equation}
S_x(\Omega)\leq A(\Omega_0)W{\Delta\Omega^2\over\Omega_0^2}+{B(\Omega_0)\over W}.
\label{eq:spdens2}
\end{equation}
Minimizing the right hand side  of the above equation with respect to $W$, we find the expression for
the  minimum noise in a fixed bandwidth:
\begin{eqnarray}
{S_x}_{optimal}(\Omega)&\leq& 2\sqrt{AB}{\Delta\Omega\over\Omega_0}=\label{line1}\\
                     &=&{1\over 2\pi^2} {\lambda^2\tau\over N}{\Delta\Omega\over \Omega_0}\sim
                        {S_x}_{\rm SQL}(\Omega_0){\Delta\Omega\over\Omega_0},\nonumber
\label{eq:spdens3}
\end{eqnarray}
which is achieved at the input power
\begin{equation}
W_{\rm optimal}={V^2\over 32\pi^2{\chi^{(3)}}^2 N\tau}\left({\Omega_0\over \omega_{\rm opt}}\right)^2
                \left({L\over l}\right)^2{1\over\hbar\omega_e} {\Omega_0\over\Delta\Omega}.
\label{eq:optpower}
\end{equation}
In the above expressions $\lambda$ is the wavelength of light inside the FP resonator
and $S_{\rm SQL}(\Omega_0)$ is the Standard Quantum Limit noise at the frequency $\Omega_0$
for a free mass. Clearly, the signal-to-noise ratio for this narrow-band measurement
is $\sqrt{{\Omega_0/\Delta\Omega}}$ greater than that in the case of the SQL :
\begin{equation}
\left({S\over N}\right)\sim\left({S\over N}\right)_{\rm SQL}\sqrt{\Omega_0\over\Delta\Omega}.
\label{eq:stn}
\end{equation}

\section{ perturbation theory for Fabry-Perot cavity with moving
         walls}

In this appendix we derive a formal series for the optical field inside a 
Fabry-Perot resonator which is pumped by a monochromatic laser beam and the walls of 
which are free to perform motions small compared to the wavelength
of light $\lambda$. The expansion parameter is $\delta x/\lambda$,
where $\delta x$ is the change of  length of the resonator.
For our purposes we are only  interested in expanding up to 
$(\delta x/\lambda)^2$; and we use this formal series to derive 
Eq. (\ref{eq:noisec}).

The following situation is considered: for simplicity we assume that light
is pumped on resonance by a laser beam $E_{\rm in}=\alpha e^{-\imath(\omega_{\rm opt} t-k x)}$
through the left mirror which is at rest and has reflectivity $r$ and 
transmissivity $T$. For concreteness it is assumed that
the fluctuations in length $\delta x$ originate  from the 
motion of the right mirror which is assumed to be perfectly reflecting.
Further, we assume that the plain wave approximation is applicable 
and hence the optical field inside the
resonator satisfies the one-dimensional wave equation:
\begin{equation}
\left({\partial^2\over \partial t ^2}-c^2{\partial^2\over\partial x^2}
\right) A(x,t)=0
\end{equation}
The general solution of  of the above equation is 
\begin{equation}
A(x,t)=f\left( t+{x\over c}\right)+g\left( t-{x\over c}\right)
\label{eq:amplitude}
\end{equation}
where $f$ and $g$ are arbitrary functions. The boundary conditions at
the left mirror ($x=0$) and at the right mirror ($x=L+\delta x$) read 
respectively
\begin{equation}
g(t)-rf(t)=T\alpha e^{-\imath\omega_{\rm opt} t}
\label{eq:boundary}
\end{equation}
and
\begin{equation}
f\left(t+\tau_0+{\delta x\over c}\right)+g\left(t-\tau_0-{\delta x\over c}\right)=0
\end{equation}
where $\tau_0=L/c$.
Eliminating $g(t)$ from these two equations, we get
\begin{eqnarray}
f\left(t+\tau_0+{\delta x\over c}\right)&-&r f\left(t-\tau_0 -
{\delta x\over c}\right)\label{line1}\\
 &=&T\alpha e^{-\imath \omega_{\rm opt} t} e^{\imath
\omega_{\rm opt}\tau_0}e^{\imath{\omega_{\rm opt}\over c}\delta x}\nonumber
\end{eqnarray}
or, expanding in $\delta x$ up to second order, 
\begin{eqnarray}
f(t+\tau_0)-r f(t-\tau_0)&=&T\alpha e^{-\imath\omega_{\rm opt} t} e^{\imath
\omega_{\rm opt}\tau_0}\left(1+\imath{\omega_{\rm opt}\over c}\delta x-{\omega_{\rm opt}^2
\over c^2}\delta x^2\right)\nonumber\\
                         & &-{1\over c}\left[f^{\prime}\left(t+\tau_0\right)+
rf^{\prime}\left(t-\tau_0\right)\right]\delta x\label{line1}\\
                         & &-{1\over 2c^2}\left[
f^{\prime\prime}\left(t+\tau_0\right)-rf^{\prime\prime}\left(t-\tau_0
\right)\right]\delta x^2\nonumber
\end{eqnarray}
We take a Fourier tranform of the above equation and then solve it by 
iterations:
\begin{equation}
f(\omega)=f^{(0)}(\omega)+f^{(1)}(\omega)+f^{(2)}(\omega)+...
\end{equation}
where
\begin{equation}
f^{(0)}(\omega)={T\alpha\over 1-r}\delta (\omega-\omega_{\rm opt}),
\end{equation}
\begin{equation}
f^{(1)}(\omega)={2\imath\omega_{\rm opt} T\alpha\over
c\left[\left(1-r\right)\cos{\left(\omega\tau_0\right)}
-\imath\left(1+r\right)\sin\left(\omega\tau_0\right)\right] (1-r)}
\delta x(\omega-\omega_{\rm opt}),
\end{equation}
\begin{eqnarray}
f^{(2)}(\omega)&=&-{4\omega_{\rm opt}^2 T\alpha\over c^2\left[\left(1-r\right)
\cos\left(\omega\tau_0\right)-\imath (1+r)\sin\left(\omega\tau_0\right)
\right] (1-r)}\nonumber\\ 
               & &\times\int d\omega^{\prime}{\cos(\omega^{\prime}\tau_0) 
\delta x(\omega^{\prime}-\omega_{\rm opt})\delta x(\omega-\omega^{\prime})
\over (1-r)\cos(\omega^{\prime}\tau_0)-\imath (1+r)\sin (\omega^{\prime}
\tau_0)}.\label{line1}
\end{eqnarray}
When writing down the above terms we took into account the fact that
$1-r\ll 1$. The structure of $f^{(2)}$ is clear: it corresponds to upconversion
of light at frequency $\omega_{\rm opt}$ to an 
intermediate frequency $\omega^{\prime}$ and then from $\omega^{\prime}$
to $\omega$, with $\omega^{\prime}$ being integrated over.
From Eqs. (\ref{eq:amplitude}) and (\ref{eq:boundary}) 
\begin{equation}
A(x,\omega)\simeq -2\imath\sin\left({\omega\over c} x\right) f(\omega).
\end{equation}

The 	 BK readout system detects the square of the amplitude of the optical field:
\begin{eqnarray}
S(x,t)&\equiv &|A(x,t)|^2\label{line1}\\
             &=&S^{(0)}(x,t)+S^{(1)}(x,t)+S^{(2)}(x,t)+...\nonumber
\end{eqnarray}
where
\begin{equation}
S^{(0)}(x,\Omega)=4\sin^2\left({\omega_{\rm opt}\over c} x\right) C^2 \delta(\Omega),
\end{equation}
\begin{equation}
S^{(1)}(x, \Omega)=0
\end{equation}
and
\begin{eqnarray}
S^{(2)}(x,\Omega)&=&-\left({2\omega_{\rm opt} C\over c}\right)^2 \int \left\{ {\sin\left[{\omega_{\rm opt}-\Omega^{\prime}\over c}x\right]
                    \sin\left[{\omega_{\rm opt}+\Omega-\Omega^{\prime}\over c} x\right]\over {\cal L} (\Omega^{\prime})
                    {\cal L} (\Omega-\Omega^{\prime})}+\right.\label{eq:s2}\\
                 & &\left. {2\left[\sin\left({\omega_{\rm opt}+\Omega\over c}x\right)+\sin\left({\omega_{\rm opt}-\Omega\over c}x\right)\right]\over
                    {\cal L}(\Omega){\cal L}(\Omega^{\prime})} \right\} \delta x(\Omega^{\prime})\delta x(\Omega-
                    \Omega^{\prime}) d\Omega^{\prime}.\nonumber
\end{eqnarray}
In the above expression $C\equiv T\alpha/(1-r)$ and \\
${\cal L}(\Omega)\equiv (1-r)\cos(\Omega\tau)+(1+r)\sin(\Omega\tau)$.

In real interferometers $\delta x$ represents, for example, motion of the 
surface of the mirror due to the thermal excitation of the test mass's internal modes.
In what follows the contribution from  the internal  mode of  lowest 
frequency is considered and then it will be shown that the sum of contributions
of all the higher modes will have the same order of magnitude. It is assumed that the thermal noise is a Markoff Gaussian
process, and therefore is described by the following equation:
\begin{equation}
\delta x(\Omega)={F(\Omega)\over \Omega^2-\Omega_m^2+\imath\gamma_m\Omega},
\label{eq:langev}
\end{equation}
where $\Omega_m$ is the eigenfrequency of the mechanical mode, $\gamma_m$ is the damping rate and $F(\Omega)$ is the Langevin
force satisfying
\begin{equation}
\langle F(\Omega_1) F(\Omega_2)\rangle ={D\over 2\pi}\delta (\Omega_1+\Omega_2),
\end{equation}
where $D=k_B T_e\gamma_m/ m^*$ is the velocity diffusion rate. Here $m^*$ is the effective mass of the mode 
(approximately given by the mirror mass $m$),
$k_B$ is  Boltzmann's constant and $T_e$ is the temperature of the enviroment. To calculate the spectral density of 
the fluctuations of $S^{(2)}$ (the goal of this analysis) we will need the 4-point correlation function :
\begin{eqnarray}
\langle F(\Omega_1) F(\Omega_2) F(\Omega_3) F(\Omega_4)\rangle &=& {D^2\over 8\pi^2}\left[\delta(\Omega_1+\Omega_2)\delta(\Omega_3+\Omega_4)\right.
                                                     \label{line1}\\ 
                                                 & &\left. +\delta(\Omega_1+\Omega_3)\delta(\Omega_2+\Omega_4)+
                                                     \delta(\Omega_1+\Omega_4)\delta(\Omega_2+\Omega_3)\right].\nonumber
\end{eqnarray}
Using the above expression, Eq. (\ref{eq:s2}) and Eq. (\ref{eq:langev}) we obtain
\begin{equation}
\langle S^{(2)}(x,\Omega_1)S^{(2)}(x,\Omega_2)\rangle =(2\omega_0 C)^4 \left[M_1\delta\left(\Omega_1\right)\delta\left(\Omega_2\right)+
                                         M_2\left(x,\Omega_1\right)\delta\left(\Omega_1+\Omega_2\right)\right].
\end{equation}
Here $M_2(\Omega)$, which characterizes the spectral density of fluctuations of $S^{(2)}$, is given by
\begin{eqnarray}
M_2 (x,\Omega)&=&{D^2\over 8\pi^2}\int \left[K\left(\Omega, \Omega^{\prime}\right) K\left(-\Omega,-\Omega^{\prime}
                   \right)+\right.\label{eq:M}\\
           & & \left. K\left(\Omega,\Omega^{\prime}\right)K\left(-\Omega,-\Omega+\Omega^{\prime}\right)\right] d\Omega^{\prime},
               \nonumber
\end{eqnarray}
where
\begin{eqnarray}
K(\Omega,\Omega^{\prime})&=&{1\over {\cal L}(\Omega^{\prime})({\Omega^{\prime}}^2-\Omega_m^2+\imath\gamma_m\Omega^{\prime})
                            \left[\left(\Omega-\Omega^{\prime}\right)^2-\Omega_m^2+\imath\gamma_m\left(\Omega-\Omega^{\prime}
                            \right)\right]}\nonumber\\
                         & &\left\{ {\sin\left({\omega_{\rm opt}-\Omega^{\prime}\over c}x\right)\sin\left({\omega_{\rm opt}+\Omega-
                            \Omega^{\prime}\over c}x\right)\over {\cal L}(\Omega-\Omega^{\prime})}\right.\label{line1}\\
                         & &\left. +{\left[\sin\left({\omega_{\rm opt}+\Omega\over c}x\right)+\sin\left({\omega_{\rm opt}-\Omega\over c}x\right)
                            \right]\sin\left({\omega_{\rm opt}\over c}x\right)\over {\cal L}(\Omega)}\right\}.\nonumber
\end{eqnarray}

It is possible to integrate Eq. (\ref{eq:M})
 exactly, but it is clear that the main contribution will come from mechanical and 
optical resonances, $\Omega^{\prime}=\Omega_{\rm opt}$ and $\Omega^{\prime}=\Omega_m$.  For $\gamma_m\ll (1-r)/\tau$ (which is the case
for, e.g., fused silica) the major contribution in Eq. (\ref{eq:M}) is due to the mechanical resonances:
\begin{equation}
M_2(x,\Omega)\sim {D^2\over 8\pi^2}(2\omega_{\rm opt} C)^2\left[K_1\left(x,\Omega\right)+K_2\left(x,\Omega\right)+K_3\left(x,\Omega\right)
\right]
\end{equation}
where
\begin{equation}
K_1(x,\Omega)\sim{2\sin^2\left({\omega_{\rm opt}-\Omega_m\over c}x\right)\sin^2\left({\omega_{\rm opt}+\Omega_m\over c}x\right)+
                     \sin^4\left({\omega_{\rm opt}+\Omega_m\over c}x\right)+\sin^4\left({\omega_{\rm opt}-\Omega_m\over c}x\right)\over
                     16\sin^4(\Omega_m\tau)\Omega_m^4\Omega^2\gamma_m},
\end{equation}
\begin{equation} 
K_2(x,\Omega)\sim{\left[\sin\left({\omega_{\rm opt}+\Omega\over c}x\right)+\sin\left({\omega_{\rm opt}-\Omega\over c}x\right)
                     \right]^2\over 16 \sin^2(\Omega_m\tau)(\Omega\tau)^2\Omega_m^4\Omega^2\gamma_m},
\end{equation}
\begin{equation}
K_3(x,\Omega)\sim{\left[\sin^2\left({\omega_{\rm opt}+\Omega_m\over c}x\right)+\sin^2\left({\omega_{\rm opt}-\Omega_m\over c}x\right)\right]
                     \sin^2({\omega_{rm opt}\over c}x)\over 16\sin^3(\Omega_m\tau)(\Omega\tau)\Omega_m^4\Omega^2\gamma_m}.
\end{equation}
We are only interested in detection frequencies such that $\Omega\tau\ll 1$. Then for the configuration of  nonlinear slabs 
shown in Fig. 2 the main contribution to the noise in the BK meter readout will come from $K_2$ and $K_3$. The spectral
density of the displacement noise will be 
\begin{equation}
S_{\delta x}\sim{1\over 8\pi^2}\left({k_B T_e\over m^*}\right)^2{\omega_{\rm opt}^2\over c^2}{\gamma_m\over\Omega_m^4\Omega^2
                         (\Omega\tau)^2}(\Omega\tau+\epsilon),
\label{eq:spdensity}
\end{equation}
where  $\epsilon$ characterizes the degree of positioning
error and the mismatch of nonlinearities of the two slabs:
\begin{equation}
\epsilon\sim\max\left({|\chi^{(3)}_1|-|\chi^{(3)}_2|\over \chi^{(3)}_1}, {\delta l\over \lambda}\right).
\end{equation}
Here $\delta l$ is the spatial offset of  the central point between  the two slabs.

Equation (\ref{eq:spdensity}) is the main result of this Appendix. It's implications are
discussed at the end of Sec. IV. 
\end{appendix}

\end{document}